\documentclass[aip, apl
 sd,%
 amsmath,amssymb,
 reprint,%
]{revtex4-1}

\usepackage{graphicx}
\usepackage{dcolumn}
\usepackage{bm}

\begin{document}

\preprint{AIP/123-QED}

\title[Hybridizing matter-wave and classical accelerometers]{Hybridizing matter-wave and classical accelerometers}

\author{J. Lautier}
\author{L. Volodimer}
\author{T. Hardin}
\author{S.Merlet}
\author{M. Lours}
\author{F. Pereira Dos Santos}
\author{A. Landragin}
 \email{arnaud.landragin@obspm.fr}
\affiliation{ 
LNE-SYRTE, Observatoire de Paris, CNRS, UPMC, 61
avenue de l'Observatoire, 75014 Paris, France
}%

\date{\today}

\begin{abstract}
We demonstrate a hybrid accelerometer that benefits from the advantages of both conventional and atomic sensors in terms of bandwidth (DC to 430 Hz) and long term stability. First, the use of a real time correction of the atom interferometer phase by the signal from the classical accelerometer enables to run it at best performances without any isolation platform. Second, a servo-lock of the DC component of the conventional sensor output signal by the atomic one realizes a hybrid sensor. This method paves the way for applications in geophysics and in inertial navigation as it overcomes the main limitation of atomic accelerometers, namely the dead times between consecutive measurements.

%
\end{abstract}

\pacs{Valid PACS appear here}
\maketitle

Atom interferometers have demonstrated to be both very sensitive and accurate inertial sensors, which opens wide area of applications in geophysics and inertial guidance. These possibilities are already exploited in absolute cold atom gravimetry~\cite{Peters}, where record sensitivities have been demonstrated~\cite{Hu,Gillot}. In the mean time, developments have been conducted in order to simplify and reduce the size of such devices to be compliant with these applications~\cite{Bodart,Bidel,Barrett}. One of the main remaining difficulties comes from the intrinsic low frequency sampling, in the order of 1 Hz, of the vibration noise in cold atom devices. This is associated to their sequential operation leading to dead times between consecutive measurements and aliasing effect of the high frequency vibration noise~\cite{Jekeli}. Different methods have been developed to overcome these difficulties. One consists in using an active or a passive isolation platform~\cite{Peters,Hu,LeGouet,Hauth}, in which the cut-off frequency is below the cycling frequency. However, these set-ups are bulky, and thus ill-suited for operation in noisy or mobile environment because of their low frequency resonances. A second method consists in increasing the cycling frequency (330 Hz has been demonstrated~\cite{McGui}) but to the price of a drastic reduction in the sensitivity, which scales quadratically with the interrogation time. A last method is based on the post-correlations between simultaneous measurements from classical and atom accelerometers~\cite{LeGouet}. It enables reaching the state of the art performance in absolute gravimetry~\cite{Louchet}, and operating an atom interferometer in a noisy environment such as a plane~\cite{Geiger}. However, in the latter, this method leads to a reduction of the effective bandwidth, and does not solve the dead time issue. Hybridizing classical and matter-wave inertial sensors enables to overcome these limitations.\\
In this paper, we demonstrate how such a combination between a force balanced accelerometer and an atomic gravimeter enables operating both instruments at maximum performances. The combined signal associates the large bandwidth of the mechanical sensor and the long term stability and the accuracy of the atomic sensor. We exploit the correlation between the acceleration measurements in order to real-time compensate the atomic interferometer phase fluctuations due to vibrations, and to operate this instrument at its maximum sensitivity. Furthermore, we take the full advantage of these correlations to correct for the drift of the mechanical accelerometer. In a reciprocal fashion, the specific advantages of one sensor are used to overtake the characteristic limitations of the other. This results in a hybrid sensor, which combines large bandwidth (DC to 430 Hz), and the bias stability and the accuracy of the atomic interferometer.\\
We work with a Mach-Zehnder-like $\pi/2-\pi-\pi/2$ gravimeter relying on stimulated Raman transitions~\cite{Kasevich1}. The sensor head of our instrument is described in~\cite{LeGouet}. The laser system and a typical measurement sequence are presented in~\cite{Merlet2}. A vertical laser beam and its retro-reflection on a mirror form the two counter-propagating Raman beams. These are used to coherently split, deflect, and recombine a cloud of cold $^{87}$Rb atoms in free-fall in the gravity field. With this geometry, the atomic phase-shift at the output of the interferometer is given by~\cite{Borde}: $\triangle$$\Phi$=$\phi_1$$-$2$\phi_2$$+$$\phi_3$=$-$$\vec{k}_\text{eff}$$\vec{g}$$T^2$ where $\phi_i$ is the phase difference between the two Raman lasers, at the position of the center of mass of the wavepacket, at the time of the i-th Raman pulse. $\vec{k}_\text{eff}$ is the effective wave-vector, $\vec{g}$ the acceleration of the Earth gravity, and $T$ the free-evolution time between two consecutive pulses. Such atomic accelerometers are thus sensitive to the relative acceleration between the free-falling atoms and the retro-reflecting mirror, that sets the phase reference for the Raman lasers. Residual vibrations of the latter thus induce a phase noise $\phi_{vib}$. Besides, the free-fall of the atoms induces a Doppler shift which requires to sweep phase-continuously the frequency difference between the two Raman lasers to drive the two-photon transitions over the interrogation time, according to $\triangle$$\omega$(t)=$\triangle$$\omega$(0)+$\alpha$t. This adds an additional contribution $\alpha$$T^2$ to the atomic phase shift $\triangle$$\Phi$. \\
Our observable is the transition probability \textit{P}. This quantity, which is derived from the measurement of the populations $N_i$ in the two output ports of the interferometer, gives access to the phase: 
\begin{equation} 
\label{Phi1}
  P=\frac{N_1}{N_1+N_2}=A-\frac{C}{2}\cos((\vec{k}_\text{eff} \vec{g}-\alpha)T^2+\phi_{vib}+\phi_0)
\end{equation}
where $A$ is the offset of the interferogram, $C$ the contrast and $\phi_0$ an additional controlled phase shift, which we set to $\pi$/2 to operate at mid fringe, where sensitivity to phase fluctuations is maximal. The determination of the frequency chirp $\alpha_0$ that allows for an exact compensation of the Doppler shift, such as $\alpha_0=\vec{k}_\text{eff} \vec{g}$ (central fringe), gives then access to \textit{g}~\cite{Peters}. 
The high sensitivity of cold-atom interferometers is such, that typical urban vibrations induce atomic phase shifts that are often greater than $\pi$. This scatters the measurement points away from mid-fringe over several interference fringes. This ambiguity in the fringe number can be lifted using an additional measurement with a conventional sensor, using for instance the post-correlation technique~\cite{Merlet1}. In the latter case however, the short term sensitivity of the measurement is not optimal, as measurement performed at the top and bottom of the fringes have low sensitivity to phase fluctuations.

By contrast, in our method, which also exploits such a correlation, we pre-compensate vibrations atomic induced phase fluctuations on the phase difference of the Raman lasers before the wavepackets are recombined, in a so-called \textit{real-time} way. This keeps the interferometer operating at mid fringe. In our work, the classical sensor is a mechanical accelerometer~\cite{Titan}. It is fixed to the retro-reflecting mirror. After being amplified by a low-noise pre-amplifier (LNA), its signal is sampled by an analog-to-digital converter (ADC). The LNA also features adjustable high-pass and low-pass filters, which can be used to remove the DC-component of the accelerometer, and add an extra anti-aliasing filter. A FPGA-based calculator weights the digitized conventional acceleration signal by the time-domain transfer function of the gravimeter~\cite{Cheinet}, which provides an estimate $\phi_{vib}'$ of $\phi_{vib}$. Finally, this phase offset $\phi_{vib}'$ is substracted to the interferometer phase by controlling with the FPGA the phase register of a direct digital synthesis device (DDS) that generates the frequency chirp $\alpha$. The use of digital electronics enables to keep a record of the correction that has been applied on the optical phase difference.


\begin{figure}[h]
       \includegraphics[width=8.5cm]{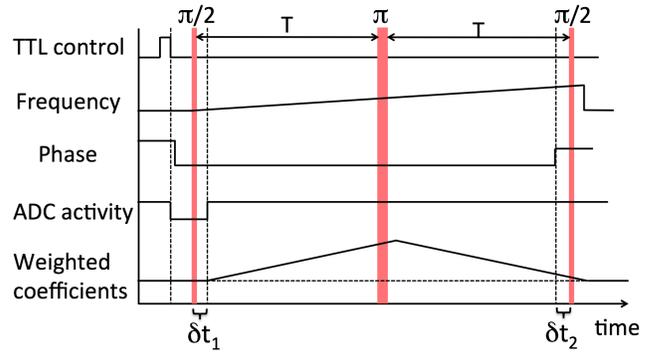}
    \caption{Chronogramm of the real-time compensation method during a measurement cycle. Our module is autonomous but operates in a synchronous manner with the interferometer.}
    \label{chrono}
\end{figure}

At each cycle, a TTL signal is delivered before the first Raman pulse by the gravimeter control system to our real-time compensation module (Fig.~\ref{chrono}). This triggers the frequency chirp and sets the phase of the DDS output signal back to a nominal value. It synchronizes the acquisition of the conventional accelerometer signal by the ADC with the time sequence of the interferometer. In order to compensate for the phase lag featured by the transfer function of the accelerometer and additional contributions of the electronics, we introduce a pure time delay $\delta t_1=1.2$~ms in the application of the sensitivity function~\cite{Merlet1}. During the interferometer, the digitized classical acceleration values are weighted in the FPGA processor by the triangle shaped transfer function of the interferometer. About $\delta t_2 = 400$~$\mu$s before the third Raman pulse, the FPGA processor ends up the calculation of the phase shift. The resulting phase correction value is then written onto the phase register of the DDS. In practice, $\phi_{vib}$' slightly differs from $\phi_{vib}$ due to the non-perfect correlations and the truncation of the triangle that amounts to 1.6~ms, discarding part of the information.



\begin{figure}[h]
    \includegraphics[width=8.5cm]{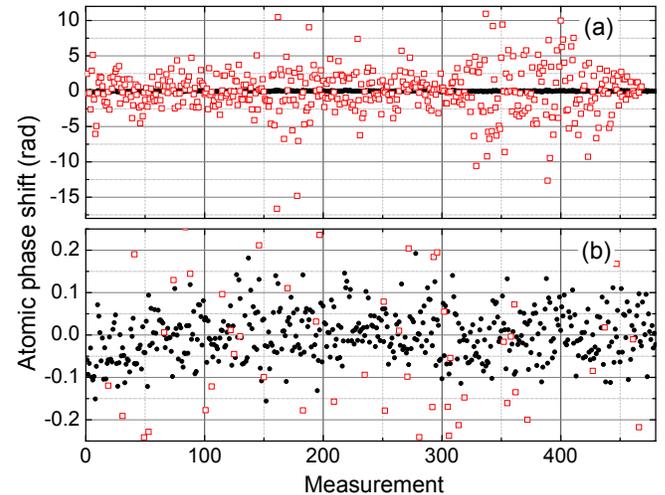}
    \caption{(color online) Effect of the real-time compensation of the vibration noise on the atomic phase. Opened red squares: atomic phase shifts estimated from the classical accelerometer. Full black points: atomic phase shifts measurements when compensated for vibrations in real time. (a) full scale and (b) zoom on the vertical axis.}
    \label{phase}
\end{figure}

\begin{figure}[h]
       \includegraphics[width=8.5cm]{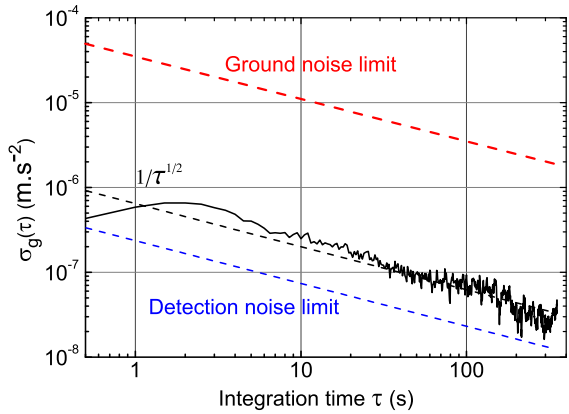}
    \caption{(color online) Allan standard deviation of the acceleration sensitivity of the gravimeter with the real-time compensation technique (RTC) without isolation plaform. The $\tau^{-1/2}$ slope represents the averaging of a white noise.}
    \label{dtegche}
\end{figure}

Fig.~\ref{phase} displays the benefit of our real-time compensation (RTC) method on the dynamics of the atomic measurement. It was performed day time in the center of Paris city, with a bandpass filter which cut-off frequencies $f_\text{lp}$ = 0.03~Hz and $f_\text{hp}$ = 10~kHz, $2T$ = 117~ms and the cycling time is $T_\text{c}$ = 500~ms. The red dots represent the calculated phase shifts $\phi_\text{vib}$' of standard deviation $\sigma_{vib}$'~=~3.3~rad that the ground vibrations should have induced on the atomic phase without any correction. The black dots stand for the residual atomic phase fluctuations with the RTC ($\sigma_{res}$~=~57~mrad). This residue is enlarged on Fig.~\ref{phase}(b). The contribution to the interferometer phase of the discarded end of the signal was calculated to remain negligible, of standard deviation on the order of 2 mrad. Nevertheless, the interferometer phase can in principle be corrected from this contribution a posteriori if necessary.
Fig.~\ref{dtegche} shows the Allan standard deviation of the measurement of \textit{g} delivered by our gravimeter with the RTC technique. The mean atomic phase is modulated through the RTC module to measure alternatively on the right and on the left of the central fringe so as to reject detection fluctuations~\cite{Louchet}. We associate to our measurement a short-term sensitivity of $6.5\times10^{-7}$~m.s$^{-2}$ at one second measurement time. The sensitivity improves up to 300 s to reach a level of $3\times10^{-8}$~m.s$^{-2}$. Compared to the level of ground vibration noise, we find a rejection efficiency of about 60. This performance is a factor of 3 above the limit arising from detection noise. 

To take full benefit of the correlation, we perform a complete hybridation of the two sensors by tying the DC component of the accelerometer signal to the  gravity changes measured by the atom gravimeter. Indeed, the DC component for most conventional accelerometers suffers from drift and temperature dependence which degrades the long term stability of the acceleration measurement. We use the measurement protocol described above to detect fluctuations of the mean atomic phase caused by drifts of the conventional sensor. From the successive values of $P$, we derive voltage corrections that we add to the signal of the auxiliary sensor output signal. This realizes an integrator loop that servo locks its DC component onto a value that corresponds to the measurement made with the atom interferometer. This hybridizes the matter-wave and classical sensors, supplying a continuous measurement of the acceleration being regularly calibrated by the absolute acceleration provided by the atom interferometer.

\begin{figure}[t]
    \includegraphics[width=8.5cm]{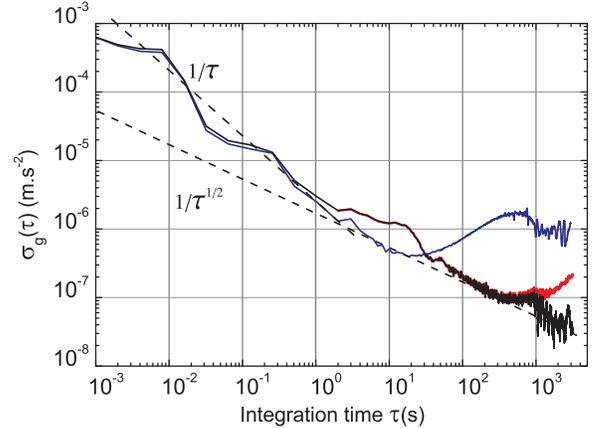}
    \caption{Allan standard deviation of acceleration signals: from the conventional accelerometer alone (blue line) and from the hybrid accelerometer without (red line) and with correction from Earth's tides (black line). }
    \label{avar}
\end{figure}

Fig.~\ref{avar} shows the Allan standard deviation of the hybrid acceleration signal, acquired by a separate acquisition unit. For this demonstration, the sensor is placed on a passive isolation platform. We compare the free-running evolution of the accelerometer signal in an open-loop configuration (in blue), with the hybrid signal (in red), eventually corrected from the Earth's tides (in black). The time constant of the integrator loop has been chosen to be about 10 s, where the sensitivities of both sensors are similar under our specific experimental conditions. For short integration time, the stability of the acceleration measurement improves in $\tau^{-1}$ as we average the high frequency vibrations without dead time. At long integration time, the stability improves in $\tau^{-1/2}$ with a sensitivity corresponding to the one of the atomic sensor when corrected from vibrations. The small bump in the stability of the hybrid sensor at 10 s is due to the servo-lock time constant. The resulting hybrid sensor features strong analogies with atomic clocks. A macroscopic device based on a classical technology is used to probe an absolute atomic reference. Then, the absolute measurement is used to servo the output signal of the local sensor, providing a continuous accurate and stable signal. The short-term stability of the hybrid sensor is determined by the self-noise of the conventional sensor. The long-term stability is set by the measurement of the atomic mean phase provided by the interferometer.

\begin{figure}[h]
    \includegraphics[width=8.5cm]{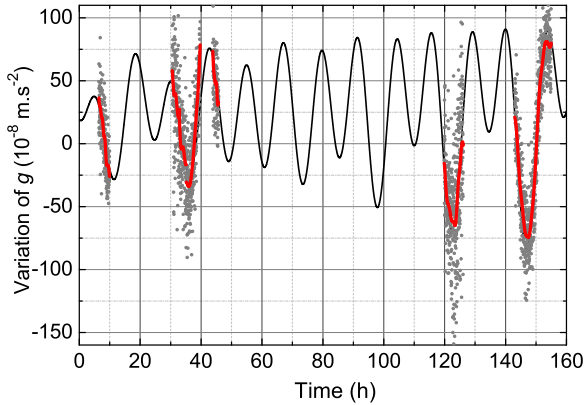}
    \caption{(color online) Long-term variations of the hybrid acceleration signal. A grey point represents 1-min averaging of the hybrid acceleration signal. The red curve is a 50-point adjacent averaging of the grey data and the thin black line is the tidal model.}
    \label{gravity}
\end{figure}

Even if we use conventional analog electronics subject to temperature fluctuations, the servo-lock on the gravimeter response makes the hybrid sensor signal offset follow the temporal variations of \textit{g} due to luni-solar tides. Fig.~\ref{gravity} shows the monitoring of the mean hybrid acceleration over 6 days. We demonstrate a good agreement of the hybrid signal with the results given by the tidal model.

In conclusion, we have demonstrated the interest of hybridizing a conventional and an atomic sensor. By using the signal of a classical accelerometer to compensate in real time the phase shift of the atom interferometer, we demonstrate that the atomic accelerometer runs at best performances without any isolation platform. In a second step, the signal of the atomic sensor is used to servo-lock the output of the conventional sensor in order to suppress its drifts and to follow gravity changes, which enables to exploit a classical accelerometer in a fluctuating thermal environment. This hybrid sensor combines the advantages of both sensors, providing a continuous and broadband (DC to 430~Hz) signal that  benefits from the long term stability and accuracy of the atomic sensor. 

These features, which make it appealing for geoscience (sismology and gravimetry), are especially of major relevance in inertial navigation. Indeed, the high frequency variation of the acceleration is the signal of interest for the calculation of the trajectory of the carrier. The loss of information from dead times is then a major limitation, which we overcome with our method. We determine that the residual bias corresponding to the accuracy of our atomic gravimeter is then the most significant contribution to the uncertainty in position. A bias of $5\times10^{-8}$~m.s$^{-2}$ leads to an error of 5 m after 4h of navigation. After a calibration stage to remove the acceleration offset, our hybrid accelerometer would allow to reach an error of less than 1 m after 4 h of navigation.

In addition, these methods can be extended to large range of acceleration by adjusting in real time not only the phase but also the difference of frequencies of the Raman lasers for a compensation of the doppler effect. Moreover, following the work presented in~\cite{Merlet1}, the servo-lock of the scaling factor of the conventional sensor can also be realized. Finally, the method developed here is general and can be extended to other type of sensors such as gyroscopes~\cite{Canuel,Tackmann} or gradiometers~\cite{Fixler,Rosi}.

\begin{acknowledgments}
The authors would like to thank the Institut Francilien pour la Recherche sur les Atomes Froids (IFRAF), the Agence Nationale pour la Recherche (ANR contract ANR - 09 - BLAN - 0026 - 01) in the frame of the MiniAtom collaboration, GRAM and CNES for funding this work. J.L. would like to thank UPMC and Coll\`{e}ge des Ing\'{e}nieurs, in the frame of the program "Science and Management" for supporting his work.
\end{acknowledgments}


\begin{thebibliography}{10}

\bibitem{Peters} A. Peters, K. Y. Chung, S. Chu, Nature, \textbf{400}, 849-852 (1999).

\bibitem{Hu} Z.K Hu, B.L. Sun, X.C. Duan, M.K. Zhou, L.L. Chen, S. Zhan, and J. Luo, Phys. Rev. A \textbf{88}, 043610 (2013).

\bibitem{Gillot} P. Gillot, O. Francis, A. Landragin, F. Pereira Dos Santos, S. Merlet, Metrologia \textbf{51}, L15-L17 (2014).

\bibitem{Bodart} Q. Bodart, S. Merlet, N. Malossi, F. Pereira dos Santos, P. Bouyer, and A. Landragin, Applied Physics Letters \textbf{96}, 134101 (2010).

\bibitem{Bidel} Y. Bidel, O. Carraz, R. Charri\'{e}re, M. Cadoret, N. Zahzam and A. Bresson, Applied Physics Letters \textbf{102}, 144107 (2013).

\bibitem{Barrett} B. Barrett, P.-A. Gominet, E. Cantin, L. Antoni-Micollier, A. Bertoldi, B. Battelier, P. Bouyer, J. Lautier, and A. Landragin, « Mobile and Remote Inertial Sensing with Atom Interferometers », Proceedings of the International School of Physics “Enrico Fermi”, Course 188 « Atom Interferometry », ed. by G. M. Tino and M. A. Kasevich Course 188, (IOS Press, Amsterdam and SIF, Bologna) 2014, pp. 493-556, arXiv:1311.7033..

\bibitem{Jekeli} C. Jekeli, Navigation 52, 1 (2005).

\bibitem{LeGouet} J. Le Gou\"{e}t, T. E. Mehlst\"{a}ubler, J. Kim, S. Merlet, A. Clairon, A. Landragin, F. Pereira Dos Santos, Appl. Phys. B \textbf{92}, 133-144, (2008).

\bibitem{Hauth}   M. Hauth, C. Freier, V. Schkolnik, A. Senger, M. Schmidt and A. Peters, Appl. Phys. B \textbf{113}, 49-55, (2013).


\bibitem{McGui} H. J. McGuiness, A. V. Rakholia, G. W. Biedermann, Applied Physics Letters \textbf{100}, 0111106  (2013).

\bibitem{Louchet} A. Louchet-Chauvet, T. Farah, Q. Bodart, A. Clairon, A. Landragin, S. Merlet, F. Pereira Dos Santos, New Journal of Physics \textbf{13}, 065025 (2011).

\bibitem{Geiger} R. Geiger, V. M\'{e}noret, G. Stern, N. Zahzam, P. Cheinet, B. Battelier, A. Villing, F. Moron, M. Lours, Y. Bidel, A. Bresson, A. Landragin, and P. Bouyer, Nature Communication \textbf{2}, 474 (2011).


\bibitem{Kasevich1}  M. Kasevich, and S. Chu, Phys. Rev. Lett. \textbf{67} 181 (1991).

\bibitem{Merlet2}  S. Merlet, L. Volodimer, M. Lours, F. Pereira Dos Santos, accepted for publication in Applied Physics B (2014).

\bibitem{Borde} Ch. Antoine and Ch. J.  Bord\'{e}, J. Opt. B: Quantum Semiclass. Opt. \textbf{5}, S199-S207 (2003).

\bibitem{Merlet1} S. Merlet, J. Le Gou\"{e}t, Q. Bodart, A. Landragin et F. Pereira Dos Santos, P. Rouchon, Metrologia \textbf{46}, 87-94 (2009).

\bibitem{Titan} Force balance triaxial accelerometer, model Titan from Nanometrics Inc., 250 Herzberg Road, Kanata, Ontario, Canada. All measurements have been realized in the $\pm$ 0.25 g full scale range mode. Web site: http://www.nanometrics.ca/products/titan.

\bibitem{Cheinet} P. Cheinet, B. Canuel, F. Pereira Dos Santos, A. Gauguet, F. Leduc, A. Landragin, IEEE Trans. on Instrum. Meas. \textbf{57}, 1141 (2008).

\bibitem{Canuel} B. Canuel, F. Leduc, D. Holleville, A. Gauguet, J. Fils, A. Virdis, A. Clairon, N. Dimarcq, C. J. Bord\'{e}, A. Landragin, and P. Bouyer, Phys. Rev. Lett. \textbf{97}, 010402 (2006).

\bibitem{Tackmann} G. Tackmann, P. Berg, C. Schubert, S. Abend, M. Gilowski, W. Ertmer, E.M. Rasel, New J. Phys. \textbf{14}, 015002 (2012).

\bibitem{Fixler} J. B. Fixler, G. T. Foster, J. M. McGuirk, M. A. Kasevich, Science \textbf{315}, 74-77 (2007).

\bibitem{Rosi} G. Rosi, F. Sorrentino, L. Cacciapuoti, M. Prevedelli and G. M. Tino, Nature \textbf{510},518-521 (2014).







‘


\end{thebibliography}
\end{document}